\documentclass[doublespacing]{elsart}

\journal{Chemical Physics}

\newlength{\spaltenbreite}
\usepackage{amsmath, amsfonts, amssymb}
\usepackage{graphicx}
\begin{document}

\begin{frontmatter}

\title{Superexchange tunneling conductance in molecular
wires}

\author{Elmar G. Petrov, Yevgen V. Shevchenko}
\address{Bogolyubov Institute for Theoretical
Physics, National Academy of Sciences of Ukraine,
Metrologichna Street 14-B, UA-03680 Kiev, Ukraine}
\author{Vladislav Snitsarev}
\address{Montclair State University, NJ 07043, USA}
\author{Victor V. Gorbach, Andrey V. Ragulya}
\address{ Nanotechcenter,
Krzhyzhanovsky Street 3, UA-03680 Kiev, Ukraine}
\author{Svetlana Lyubchik}
\address{ REQUIMTE, FCT, Universidade Nova de Lisboa, Quinta de Torre, 2829-516, Caparica, Portugal}

\date{\today}

\begin{abstract}

The modified superexchange model is used to derive the expression for nonresonant tunneling conductance mediated by
localized and delocalized molecular orbitals associated with the terminal and the interior molecular units respectively.
The model is shown to work as long as delocalization of electron density  in the chain's molecular orbitals is sustained during the tunneling.
The criteria for reduction of the superexchange model of charge tunneling to the flat barrier model are formulated and the parameters of the barrier model (energy gap and effective electron mass) are specified in the terms of inter-site coupling and energy distance from the Fermi level to the delocalized wire's HOMO level. Application of the theory to
the experiment shows that the modified superexchange model
is quite appropriate to explain the experimental results in case of the  nonresonance tunneling conductance  in --(CH$_2)$$_N$--NH$_2$ and HOOC--(CH$_2)$$_N$--COOH molecular wires.

PACS: 05.60.Gg, 73.63.Nm, 85.65/+h
\end{abstract}

\begin{keyword}  molecular wires;
nonresonant tunneling; conductance
\end{keyword}

\end{frontmatter}

\section{Introduction}

The use of organic molecules  is becoming one of the
 major strategies in miniaturization of  electronic, optoelectronic  and spintronic circuit components  \cite{carter82,nitz01,han02,mcc04,cs10,pat13,arad13,ram14}. A significant  progress in this direction has been achieved by applying  scanning tunneling  and  atomic force microscopes  for  monitoring and controlling  charge transfer processes in molecular junctions as well as for fabrication of molecular structures with desirable conduction properties \cite{rat13,bag15,xiwa16,bag17,elk17}. A molecular junction where single molecules or self assembled monolayers (SAMs) are embedded between the electrodes
can fulfill the functions of molecular wires, diodes, transistors, registers,  switches etc. \cite{zha15,cap15,metz15}. A number of factors such as  the molecule-electrode couplings, the energy position of molecular orbitals (MOs) relative to the Fermi-levels of the contacts, electronic density of states,  conformation mobility of the molecule etc., controls the current-voltage  and conductance characteristics of single molecules and molecular compounds. Therefore,  the efficiency of the charge transmission pathways depends strongly on the type of the molecular junction as well as  magnitude and polarity of the applied electric field.

The mechanism of formation of the tunneling conductance in the  molecular junction "metal - molecule -  metal"  where a molecule comprises a regular chain anchored to the electrodes through its terminal units is  of great importance.
These units bind  the chains to the metallic surface thus forming  the SAM of regular chains.
As part of the SAM,  each molecule  functions as molecular wire and thus can mediate  transmission of an electron/hole from one electrode to another.
Due molecular wire determining the a distant electron/hole transfer, the specification  of the factors that control the wire conductance at different regimes of charge transmission remains  the central problems in molecular electronics. One of the  working regimes is associated with  nonresonant charge tunneling  where the MOs of the molecular wire are not occupied by the transferred electron/hole. At such a regime,   both the current and the conductance decay exponentially with  molecular length \cite{xiwa16,sel02,cui02jpc,cui02,eng04,fan06,sim13,wie13,jos14}.
The analysis of conductivity/resistance  in molecular wires is mostly performed with the simple flat-barrier Simmons model \cite{simm63}.  The model predicts an exponential decrease in the tunneling current and conductance where the attenuation factor  $\beta$  is expressed via two fitting parameters, the effective mass  $m^*$  and the height of rectangular  barrier $\Delta E$. Detail analysis of the Simmons model shows \cite{sel02,cui02jpc,eng04} that the choice of the above mentioned fitting parameters, especially $\Delta E$, depends on the precise voltage region and the chain length. Thus, for molecular junctions, the rectangular barrier model  does not have the unified parameters.

The model of superexchange  tunneling through a molecular wire provides an alternative approach based on mutual overlap of wave functions of the bridging interior wire units as well as on the overlap of wave functions of the terminal wire units and the electrodes. This leads to formation of a direct distant coupling between the conductive states of the  spaced electrodes. The McConnell's version of superexchange model \cite{mcc61} was successfully used to describe a distant  hole transfer through DNA molecules \cite{jor02,bix02,tre02} as well as  combined hopping-tunneling electron transmission in the terminated molecular wires \cite{pzmh07}. McConnell model has also been  used to   analyze the  $I/V$ characteristics of alkane chains \cite{eng04,ram02}. The model explains the exponential drop of the current with the increase of the wire length, however,  it shows  discrepancy with the attenuation factor predicted by the barrier model. In the superexchange model, the attenuation factor is determined through  the hopping matrix element between the  neighboring sites of electron/hole localization in a regular chain, and  the energy distance of the Fermi level with respect to position of the
\emph{localized} MO belonging the interior wire unit. This energy distance  differs strongly on the barrier height $\Delta E$, which, in  case of molecular junction, is assumed to be the gap between the Fermi level and the \emph{delocalized} HOMO level belonging to the regular range of the wire \cite{cui02jpc,eng04}.

In this paper, the modified theory of nonresonant superexchange tunneling is used to analyze the dependence of the conductance of the terminated molecular wire on the length of the wire's  regular range. The explicit expressions for the conductance are derived along with the attenuation factor, an important parameter that describes the efficiency of the tunneling across the molecular junction.
In limiting cases, the attenuation factor  yields two  different limits corresponding   the Simmons or McConnel models.

The paper is organized as follows. In Section 2, the basic principles of the modified superexchange model are presented and distinct expressions for the conductance of  linear terminated molecular wires are derived. Results concerning the applicability of the model to description of the conductance in soecific molecular junctions are given in Section 3. Concluding remarks are presented in Section 4.

\section{Theoretical base}.

\begin{figure}
\includegraphics[width=7.3cm]{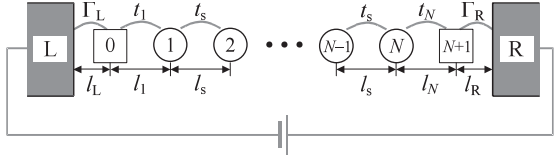}
\caption{Arrangement of  units of  a linear molecular wire relative to the attached electrodes L and R. Terminal units are denoted via $ 0$ and $ N+1$, the units of a regular chain (interior range of the wire) are $n=1,2,...N$. Quantities $\Gamma_L$  and $\Gamma_R$ are the width parameters  characterizing the broadening of respective  terminal orbital energies $E_0$ and $E_{N+1}$. Rest explanations in the text.
}
\label{fig1}
\end{figure}

We consider  molecular junction  as a quantum  system where a linear molecular wire is attached to the left (L) and the right (R) electrodes, Fig. \ref{fig1}. Bearing in mind the application of the theory to the analysis of  the tunneling conductivity in the molecular junctions, where energies of the \emph{highest occupied molecular orbitals} (HOMOs) are closer to the electrode's Fermi level  compared to the energies of the \emph{lowest unoccupied molecular orbitals} (LUMOs), only  the formation of a superexchange charge transfer with participation of  the virtual HOMOs  is  considered here.  We use the tight-binding model where  the transferred electron can leave  the twofold filled energy level of the HOMO$_n$ located on the wire unit $n = (0.1,...N,N+1)$.
The distance $l_{n n \pm 1}\equiv  l_s$ between the  neighboring units is associated with the distance  between the sites of main electron localization within the unit. For instance, in the  $N -$ alkane chain, the  $l_s$ refers to the distance between the neighboring  C--C bonds. For the sake of definiteness, let us assume that the left electrode  is grounded so that the chemical potential of the $r$th electrode appears as $\mu_{r} = E_{F} -  |e|V\delta_{r,R}$, $(r=L.R)$, where $E_F$ is the energy of electrode's Fermi level.
In the linear approximation over the bias voltage  $V = (\mu_L - \mu_R)/|e|$, the  energy of an electron  on the $n$th unit
reads $E_{n} = E^{(0)}_{n} - \eta_n |e|V$,
where  $E^{(0)}_{n}$ is the zero bias orbital energy and  $\eta_n$ is the factor that characterizes the Stark shifts
of the orbital energies.  With the grounded left electrode, this yields $\eta_{L(R)} = l_{L(R)}/l$ at $n=0(N+1)$  and $\eta_n = [l_L+l_1 + (n-1)l_s]/l$ at $n=1,2,...N$, with  $l=l_L+l_1 + (N-1)l_s + l_{N} + l_R$ being the total  interelectrode distance. The electron couplings between the MOs of the neighboring wire units are characterized by the hopping matrix elements $t_{n, n+1}$. For  the  interior (regular) part of  a molecular wire, we set  $t_{n, n+1}\equiv t_s$ whereas $t_{0, 1}\equiv t_1$ and  $t_{N, N+1}\equiv t_{N}$  are used for the terminal units, Fig. \ref{fig1}.
Interaction of the chain  with the electrodes is provided by its terminal units $n=0$ and $n = N+1$. We consider the LWR systems where orbital energies $E_{0}$ and $E_{N+1}$  differ from the interior unit energies $E_n$. Thus,  a  mixs between the MO  belonging the terminal unit and the MO of  the nearest interior wire unit is assumed to be so insignificant  that
the localization of the terminal MOs is conserved during the electron/hole transmission across the wire. As a result, the interaction between the terminal and interior units can be considered as the perturbation.  The same refers to the interaction  between the terminal MOs  and each electronic conduction state of the electrodes \cite{foot1}.

\subsection{Tunneling current and conductance}

\begin{figure}
\includegraphics[width=7.3cm]{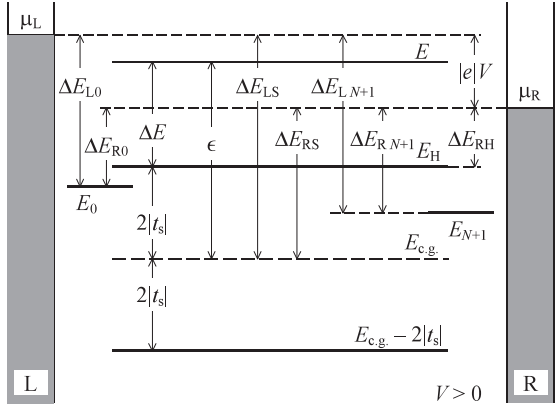}
\caption{Position of  the transmission energy $\epsilon = E - E_{c.g.}$  with respect to the "center of gravity" of  electronic density distributed over  the delocalized HOMOs.  When tunneling energy   $E$ enters in window $\mu_L\geq E\geq \mu_R$, than   the $\epsilon$ varies in range $[\Delta E_{Rs}, \Delta E_{Ls}]$.  Quantity $\Delta E_{r0(N+1)} $  is the  energy gap between chemical potential of  the $r(=L,R)$th electrode  and orbital energy  of the $0(N+1)$th terminal unit. $\Delta E = E - E_H$ is the energy distance between  tunneling energy  and the HOMO level position $E_H$ for a  long ($N\gg 1$) regular chain. In the pre-resonant tunneling regime, when $E - E_H \ll 2|t_s|$,  quantity $\Delta E $   can be referred  to the height of the apparent rectangular  barrier.
}
\label{fig2}
\end{figure}

The noted tight binding model
has been used to derive  distinct expressions for a nonresonant  tunneling current $I$ through  the terminated molecular wire. From ref. \cite{pet18},  in the framework of tight-binding model, the Landauer-B\"utteker approach \cite{dat95,tian98,muj02} gives  the following basic (integral) form for the current:
%
\begin{equation}
I = i_0\,
\,\int_{\Delta E_{Rs}}^{\Delta E_{Ls}}\,d\epsilon\,
T_L(\epsilon - \Delta E_{0s})T_{reg}(\epsilon,N) T_R(\epsilon -\Delta E_{N+1s})
\, \label{curgel0}
\end{equation}
where   $i_0\equiv (|e|/\pi\hbar)\times$1 eV $\approx 77.3 \mu$A is the current unit. In Eq. (\ref{curgel0}), the integration limits coincide with energy gaps (see also  Fig. \ref{fig2})
%
\begin{equation}
\Delta E_{rs} = \Delta E^{(0)}_{s} + |e|V\,
[\eta_{c.g.}\delta_{r,L} - (1 - \eta_{c.g.})\delta_{r,R}]\,
\label{xes}
\end{equation}
where $\Delta E_s^{(0)} = E_F - E_{s}^{(0)} > 0$ is the main transmission energy gap in an unbiased LWR.
Eq. (\ref{curgel0}) shows that the wire transmission function  is represented as  the product of three functions. Among them
%
\begin{equation}
 T_L(\epsilon - \Delta E_{0s}) = \frac{\Gamma_L}{t_s}\frac{t_1^2}{(\epsilon -\Delta E_{0s})^2 +\Gamma_L^2/4}\,
\label{ltr0}
\end{equation}
and
%
\begin{equation}
 T_R(\epsilon -\Delta E_{N+1s}) = \frac{\Gamma_R}{t_s}\frac{t_N^2}{(\epsilon -\Delta E_{N+1s})^2 +\Gamma_R^2/4}\,.
\label{rtr0}
\end{equation}
refer to  the terminal units. In Eqs. (\ref{ltr0}) and (\ref{rtr0}),
$\Gamma_L$ and $\Gamma_R$ are the width parameters that characterize  broadening of the respective terminal energies $E_0$ and $E_{N+1}$ caused by  interaction of the levels with the attached  electrodes. Quantities
%
\begin{displaymath}
\Delta E_{0s} =\Delta E_{0s}^{(0)} + |e|V(\eta_{c.g}-\eta_L)\,,
\end{displaymath}
\begin{equation}
\Delta E_{N+1s} = \Delta E_{N+1s}^{(0)} - |e|V(1-\eta_{c.g.} -\eta_R)\,.
\label{x1n0}
\end{equation}
are the energy distances $\Delta E_{0(N+1)s} = E_{0(N+1)} - E_{c.g.}$ between the terminal levels and the position of the "center of gravity" of electron density distributed over the delocalized MOs. The quantity
$\Delta E_{0(N+1)s}^{(0)}  = E_{0(N+1)}^{(0)} - E_s^{(0)} $ is the unbiased energy distance between MO's levels of the $0(N+1)$th terminal unit and the interior unit. As to the
transmission function of a regular chain (interior range of the wire) it reads
%
\begin{equation}
  T_{reg}(\epsilon,N) = \frac{\sinh^2{[\beta (\epsilon )/2]}}{\sinh^2{[(N+1)\beta(\epsilon)/2]}}\,
\label{ddfreg1}
\end{equation}
where
%
\begin{equation}
  \beta(\epsilon) = 2\ln{\big[(\epsilon/2|t_s|)+\sqrt{(\epsilon/2|t_s|)^2 - 1}\big]}\,,\;\;\;(\epsilon = E - E_{c.g.} > 0)\,,
 \label{df}
\end{equation}
is the attenuation factor per one chain unit. It
characterizes a decrease of the
$T_{reg}(\epsilon,N)$  depending on the number of chain units
$N$.  Expression (\ref{df}) exists only if the  inequality
%
\begin{equation}
 2|t_s|/\epsilon <1
\label{ets}
\end{equation}
is satisfied at the nonresonant tunneling.

In the integrand of Eq. (\ref{curgel0}), a voltage dependence is present   only in terminal transmission functions. Therefore, the  tunneling  conductance of a molecular wire,  $g = \partial I/\partial V$, appears as the sum of two contributions:
%
\begin{equation}
g = g^{(1)} + g^{(2)}\,.
\label{gg}
\end{equation}
Introducing the conductance unit $g_0 = |e|i_0 = e^2/\pi\hbar = 77.3 \mu$S, for the first contribution one obtains:
%
\begin{displaymath}
g^{(1)} = g_0\big[\eta_{c.g.}T_L(\Delta E_{L0}) T_{reg}(\Delta E_{Ls},N)T_R(\Delta E_{LN+1})
\end{displaymath}
\begin{equation}
+
(1 - \eta_{c.g.})T_L(\Delta E_{R0}) T_{reg}(\Delta E_{Rs},N)T_R(\Delta E_{RN+1}) \big]\,.
\label{g1}
\end{equation}
Here, terminal transmission functions (\ref{ltr0}) and  (\ref{rtr0}) comprise the gaps
%
\begin{displaymath}
\Delta E_{r0} = \Delta E_0^{(0)} + |e|V[\eta_L\,\delta_{r,L} - (1-\eta_L)\,\delta_{r,R}]\,,
\end{displaymath}
\begin{equation}
\Delta E_{rN+1} = \Delta E_{N+1}^{(0)} - |e|V[\eta_R\,\delta_{r,R} + (1-\eta_R)\,\delta_{r,L}]\,
\label{tergap}
\end{equation}
whereas the gaps for  chain transmission functions (\ref{ddfreg1}) are  $\Delta E_{rs}$,  Eq. (\ref{xes}). The second conductance contribution appears in the integral form:
%
\begin{displaymath}
g^{(2)} = (g_0/|e|)\int_{\Delta E_{Rs}}^{\Delta E_{Ls}}\,d\epsilon\, T_{reg}(\epsilon,N)\Big[
\frac{\partial  T_{L}(\epsilon -\Delta E_{0s})}
{\partial V} T_{R}(\epsilon -\Delta E_{N+1s})
\end{displaymath}
\begin{equation}
+ T_{L}(\epsilon -\Delta E_{0s})\frac{\partial  T_{R}(\epsilon -\Delta E_{N+1s})}{\partial V}\Big]\,.
\label{g2}
\end{equation}
The expressions for  current  and  conductance  are true for the molecular junctions   where  charge transmission is formed with participation of the localized  and delocalized HOMOs  belonging respectively to the terminal and the interior wire units. The energies of the HOMOs are represented in Fig.\ref{fig2}.
Rigorous analysis shows \cite{pet18} that the  delocalized  chain HOMOs can be involved (virtually) in formation of the superexchange tunneling  only if the inequality
%
\begin{equation}
|\Delta_s/2t_s|\, S(N) \ll 1
\label{inedel}
\end{equation}
is satisfied in the LWR system. In Eq. (\ref{inedel}),
%
\begin{equation}
\Delta_s = |e|V(l_s/l)\,
\label{ds}
\end{equation}
is the energy drop between identical neighboring units and
%
\begin{displaymath}
 S(N) =
 \Big(\frac{1}{N+1}\Big)\Big[\frac{1}{1-\cos{\big(\frac{\pi}{N+1}}\big)} - \frac{1}{1-\cos{\big(\frac{3\pi}{N+1}}\big)}\Big]
 \end{displaymath}
 \begin{equation}
 \times\Big[\frac{1}{\cos{\big(\frac{\pi}{N+1}}\big) - \cos{\big(\frac{2\pi}{N+1}}\big)}\Big] \,
\label{inedel2}
\end{equation}
is the function that depends solely on the number of chain units. If the inequality (\ref{inedel}) is satisfied,  then
energies of the delocalized HOMOs  are given by equation
%
\begin{equation}
{\mathcal E}_{\nu} = E_{c.g.} - 2|t_s|\cos{\Big(\frac{\pi\nu}{N+1}\Big)}\,,\;\;\; (\nu = 1,2,...N),
\label{endel}
\end{equation}
with
%
\begin{equation}
 E_{c.g.}  = E_s^{(0)} - |e|V\eta_{c.g.}\,
\label{shdel}
\end{equation}
being the energy position of the "center of gravity" of the electron density for the delocalized HOMOs. It should be particularly emphasized that the Stark shift is identical for each energy level related to the delocalized orbitals.

\subsection{Explicit expressions for a conductance}

Reading form for the first conductance contribution follows from Eq. (\ref{g1})  taking into account Eqs. (\ref{ltr0}), (\ref{rtr0}), (\ref{ddfreg1})  and the relation
%
\begin{equation}
\cosh{[\beta(\epsilon)/2]} = \epsilon /2|t_s|\,.
\label{et}
\end{equation}
This yields
%
\begin{displaymath}
  g^{(1)}= g_0 \Big[\eta_{c.g.}\Big(\frac{\Gamma_L\Gamma_R}{\Delta E_{Ls}^2}   \Big)\frac{t_1^2t_{N}^2\Phi(\beta_L,N)}{(\Delta E_{L0}^2 + \Gamma_L^2/4)(\Delta E_{LN+1}^2 + \Gamma_R^2/4)}
 \end{displaymath}
 \begin{equation}
 + (1-\eta_{c.g})\Big(\frac{\Gamma_L\Gamma_R}{\Delta E_{Rs}^2}   \Big)\frac{t_1^2t_{N}^2\Phi(\beta_R,N)}{(\Delta E_{R0}^2 + \Gamma_L^2/4)(\Delta E_{RN+1}^2 + \Gamma_R^2/4)}\Big]\,.
\label{g1a}
\end{equation}
Here,  $\Phi(\beta_{L(R)},N)$ is the  chain attenuation function
%
\begin{equation}
\Phi (\beta (\epsilon),N)  =
\frac{\sinh^2{\beta (\epsilon)}}{\sinh^2{[(N+1)
(\beta (\epsilon)/2)]}} \,                                                                                                                                                                                                                                                                                                                                                \label{lf}
\end{equation}
with attenuation factor  (\ref{df})  taken  at $\epsilon = \Delta E_{rs}$,  $(r=L,R)$.
Bearing in mind property $\Phi (\beta (\epsilon),1)$ =1, the function (\ref{lf}) becomes a very suitable value to characterize the superexchange tunneling drop dependence on the chain length.

To obtain  a reading form for the second conductance contribution,  Eq. (\ref{g2})   we employ the approach previously proposed  \cite{pet18} for reduction of the integral form for the current, Eq. (\ref{curgel0})  to more simple  analytic forms.   One of  them is derived  using  the so called mean-value (m.v.) approximation. This leads to a nearly identical dependence of the $I$ on $V$ and $N$ as  given by  basic integral form  (\ref{curgel0}). In our case,  in line with the m.v. approximation,
the transmission functions  $T_{L(R)}$, $T_{reg}$ and derivatives $\partial T_{L(R)}/\partial V$
are  substituted for averaged values $\overline{T}_{L(R)}$,
$ \overline{T}_{reg}(N)$ and
$\overline{\partial T_{L(R)}/\partial V }$, respectively.
This reduces  Eq. (\ref{g2}) to
%
\begin{displaymath}
 g^{(2)}\simeq g^{(2)}_{m.v.}\approx  4g_0 \Big[\frac{(\eta_{c.g} - \eta_L)(\Delta \epsilon_{0}\Gamma_L/t_s^2) t_1^2t_{N}^2}{(\Delta E_{L0}^2 + \Gamma_L^2/4)(\Delta E_{R0}^2 + \Gamma_R^2/4)} \chi_{N+1}
\end{displaymath}
\begin{equation}
- \frac{(1-\eta_{c.g} - \eta_L)(\Delta \epsilon_{N+1}\Gamma_R/t_s^2) t_1^2t_{N}^2}{(\Delta E_{LN+1}^2 + \Gamma_L^2/4)(\Delta E_{RN+1}^2 + \Gamma_R^2/4)}\chi_0\Big]\,\overline{T}_{reg}(N)
\label{mvfd}
\end{equation}
where
\begin{equation}
\chi_{0(N+1)}  = {\rm tan}^{-1}{\Big(\frac{2\Delta E_{L0(N+1)}}{\Gamma_{L(R)}}\Big)}
- {\rm tan}^{-1}{\Big(\frac{2\Delta E_{R0(N+1)}}{\Gamma_{L(R)}}\Big)}\,.
\label{xlr}
\end{equation}
With use of expressions  \cite{pet18}
%
\begin{equation}
\overline{T}_{reg}(1) = \frac{t_s^2}{\Delta\epsilon_s^2 - (|e|V/2)^2}\,,
\label{att1}
\end{equation}
%
%
\begin{displaymath}
\overline{T}_{reg}(2) = \frac{t_s^2}{4}\Big\{
\frac{1}{|e|Vt_s}\ln{\Big[\frac{\Delta\epsilon_s^2 - (t_s + |e|V/2)^2}{\Delta\epsilon_s^2 - (t_s - |e|V/2)^2}
\Big]}
\end{displaymath}
\begin{equation}
+ \Big[
\frac{1}{(\Delta\epsilon_s - t_s)^2  - (|e|V/2)^2}
 + \frac{1}{(\Delta\epsilon_s + t_s)^2  - (|e|V/2)^2}\Big]\Big\} \,.
\label{att2}
\end{equation}
and
%
\begin{displaymath}
\overline{T}_{reg}(N\geq 3)
\simeq \Big(\frac{t_s}{|e|V}\Big)\frac{1}{2N-1}
\end{displaymath}
\begin{equation}
\times
\Big[F(\beta_R){\rm e}^{- \beta_R[N-(1/2)]}
- F(\beta_L){\rm e}^{- \beta_L[N-(1/2)]} \Big]\,,
\label{trregav}
\end{equation}
where
%
\begin{equation}
F(\beta) = 1 - (2N-1)\Big[\frac{3}{2N+1}{\rm e}^{-\beta} + \frac{3}{2N+3}{\rm e}^{-2\beta}
 \frac{1}{2N+5}{\rm e}^{-3\beta}\Big]\,,
\label{f}
\end{equation}
we obtain an explicit form for the second conductance contribution. It is important to note that attenuation factors $\beta_{L}$   and $\beta_{R}$ are identical to those in Eq. (\ref{g1a}).

\section{Results and discussion}

\begin{figure}
\includegraphics[width=7.3cm]{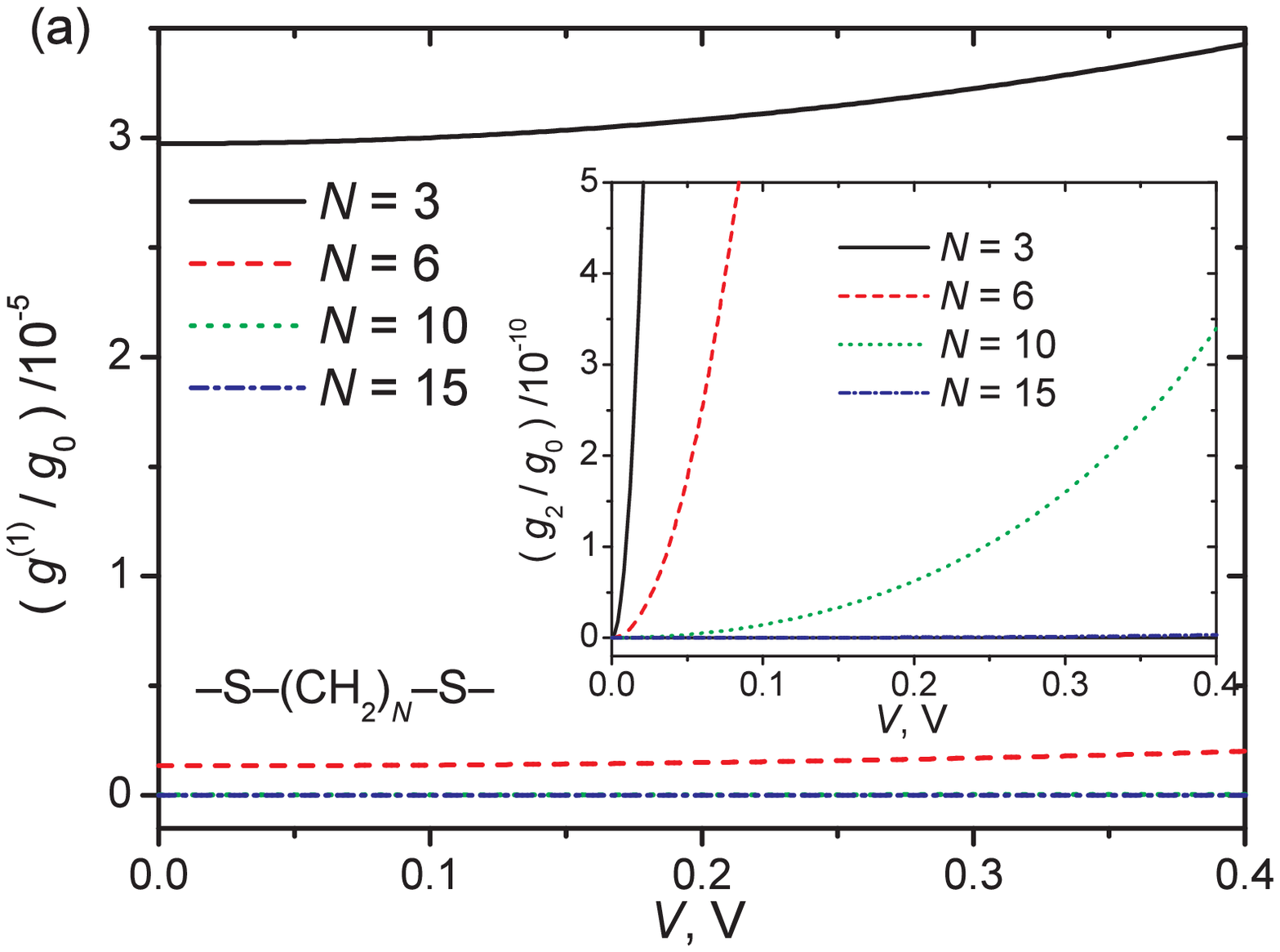}
\\

\includegraphics[width=7.3cm]{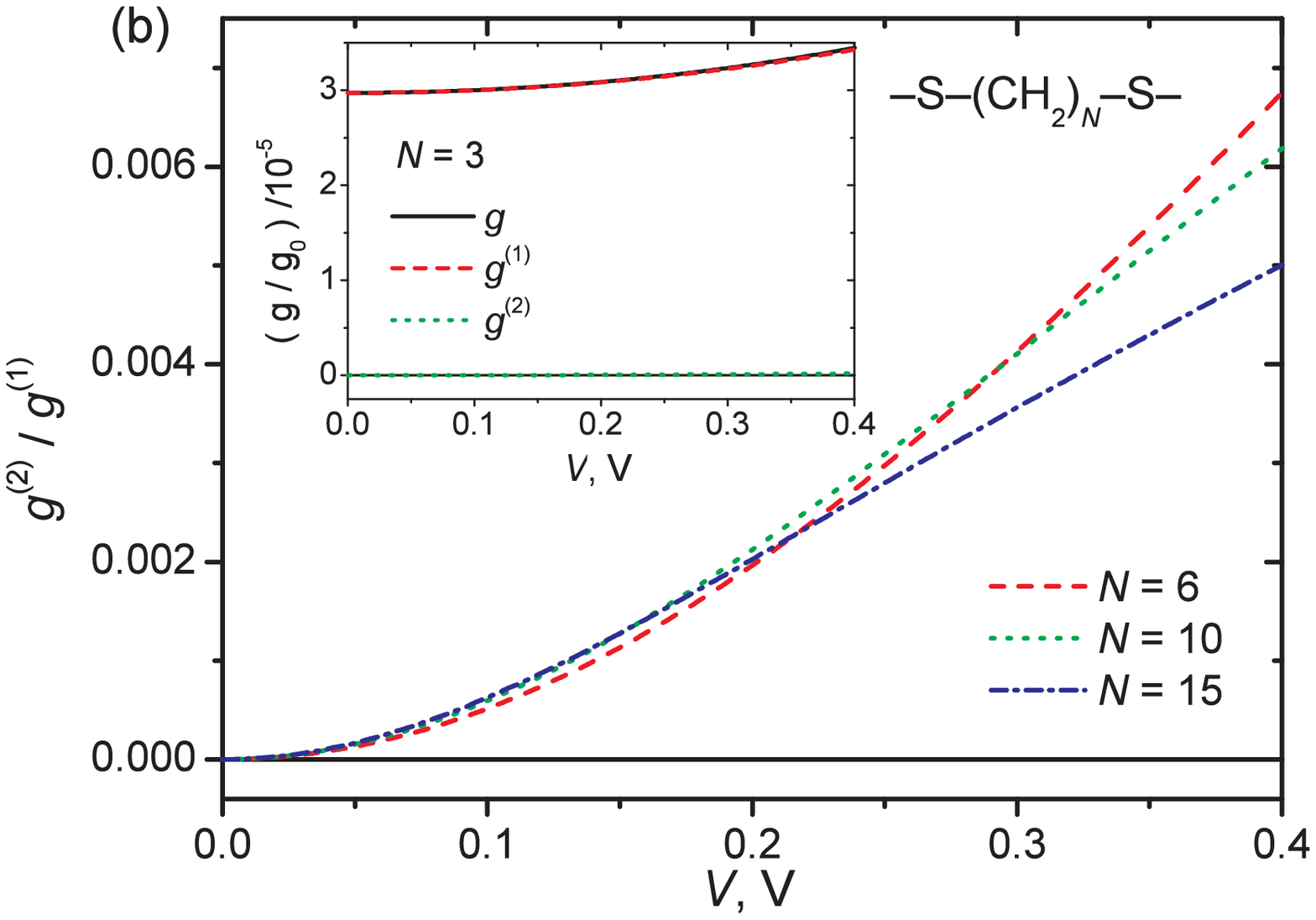}
\caption{Voltage dependence of conduction contributions $g^{(1)}$ and $g^{(2)}$ on the applied voltages (a).   Ratio of the contributions shows a minor weight of the $g^{(2)}$ in common $g$ (b). Parameters of the modified superexchange model are the same that have been utilized in ref. \cite{pet18} for the --S--(CH$_2$)$_N$--S-- wire: $\Delta E_*$ = 3.4 eV, $t_{*}$ = 2.5 eV, $\Delta E_s^{(0)}$ = 6.3 eV, $\Gamma_*$ = 0.2 eV.
}
\label{fig3}
\end{figure}

To demonstrate the mechanism of formation of the nonresonant superexchange tunneling conductance,  we consider the perfectly symmetric LWR  system where the wire  is the $N-$ alkane chain anchored  to the gold electrodes via terminal units X = --SH, --NH$_2$, --COOH.  The experimental data on high and low conductance  of  the X--(CH$_2)$$_N$--X wires
as a function of molecular length are well represented in paper  \cite{fan06}. The voltage region covers [-0.4, +0.4] V and the number of CH$_2$ groups  is changed from 2 to 12.  In such conditions, the orbital energies $E_{0} $ and $E_{N+1}$  do not enter in resonance with electrodes's Fermi levels. Besides, transmission gaps in Eqs . (\ref{g1a}) and  (\ref{mvfd})  exceed  broadenings  $\Gamma_L = \Gamma_R \equiv \Gamma_{*}$. This yields $\chi_{0(N+1)} \approx |e|V\Gamma_*/[2\Delta E_{L0(N+1)}\Delta E_{R0(N+1)}]$.
Introducing  $t_{*}\equiv t_1 =  t_{N+1}$ along with
$\eta_*\equiv \eta_L = \eta_R$ and  bearing in mind the fact that independently on the chemical structure of molecular junction the factor $\eta_{c.g.}$ is equal to 1/2, for  the first conductance contribution one obtains
%
\begin{equation}
g^{(1)}= \frac{g_0}{2}\Big[\frac{\Gamma_*^2t_*^4}{\Delta E_{Ls}^2\Delta E_{L0}^2\Delta E_{LN+1}^2}\Phi(\beta_L,N)
 + \frac{\Gamma_*^2t_*^4}{\Delta E_{Rs}^2\Delta E_{R0}^2\Delta E_{RN+1}^2}\Phi(\beta_R,N)\Big]\,.
\label{g1b}
\end{equation}
The dependence of $g$ on $N$ is concentrated in the terminal gaps (\ref{tergap}) and the attenuation functions $\Phi(\beta_{L(R)} ,N)$.
Because $\Phi(\beta_r,1) = 1$, the  function $\Phi(\beta_R,N)$
is quite suitable for characterization of conductance drop   with chain length increase. As to the second contribution, it appears as
%
\begin{displaymath}
g^{(2)}\approx g^{(2)}_{m.v.} = g_0|e|V(1-2\eta_*)  \frac{(\Gamma_*^2t_*^4/t_s^2)}{\Delta E_{L0}\Delta E_{R0}\Delta E_{LN+1}\Delta E_{RN+1}}
 \end{displaymath}
 \begin{equation}
  \times \Big[\frac{\Delta \epsilon_0}{\Delta E_{L0}\Delta E_{R0}} - \frac{\Delta \epsilon_{N+1}}{\Delta E_{LN+1}\Delta E_{RN+1}}\Big]\overline{T}_{reg}(N)\,.
\label{g1b2}
\end{equation}
To estimate the numerical weight of $g^{(1)}$ and $g^{(2)}$ in the total conductance $g$ let us refer to the results concerning the application of the  modified  superexchange model to description of  a nonresonant  tunneling current through --S--(CH$_2)$$_N$--S-- wire. To this end, let us note that the model contains two fundamental parameters, the zero bias gap $\Delta E_s^{(0)}$ and the intersite coupling $t_s$ (for  alkane chains, parameter $t_s$ is positive so that $t_s =|t_s|$).  These parameters determine the most important  wire characteristic, attenuation factor $\beta_0 \equiv  \beta (\epsilon = \Delta E_s^{(0)})$.  Strong relation between above parameters is fixed  with the basic equality (\ref{et}).     In the case of
molecular wire with X = SH, NH$_2$  and  COOH, the $\beta_0$  takes the values  1.02, 0.83 and 0.80 (per CH$_2$ group), respectively \cite{fan06}. Therefore, corresponding magnitudes  for the ratio $\Delta E_s^{(0)}/2t_s $  are 1.133, 1.087 and 1.081.   The second important relation between parameters $\Delta E_s^{(0)}$ and $2t_s$  follows from the condition at which  HOMO energy $E_H = {\mathcal E}_{\nu = N(\gg 1)} = E_{c.g.} + 2t_s$ enters in resonance with the Fermi energy of one of the electrodes. At a positive polarity, this occurs at $V = V_{cr}$ where
%
\begin{equation}
V_{cr} = \frac{2}{|e|}(\Delta E_s^{(0)} - 2t_s)\,.
\label{et2}
\end{equation}
In HS--(CH$_2)$$_N$--SH wire, the $V_{cr}$ is presumably about 1.5 V. (No conductance peaks are observed  outside of 1.5 V \cite{eng04}). Therefore,  using the expressions (\ref{et}) and (\ref{et2}), one obtains $\Delta E_s^{(0)} \approx 6.3$ eV. $t_s\approx 2.78$ eV.  These values have been used in ref. \cite{pet18} to explain the $I/V$ characteristics of  the --S--(CH$_2)$$_N$--S-- wire. Our calculations of  the contributions  $g^{(1)}$ and $g^{(2)}$, which have been presented in Fig. \ref{fig3}, show that in the case of charge tunneling across --S--(CH$_2)$$_N$--S-- wire
the $g^{(1)}$ exceeds  the $g^{(2)}$ significantly, so that $g \simeq g^{(1)}$ (cf. the insertion in Fig. \ref{fig3}b).
Physically, this result is explained by the alignment of the terminal energies $E_{0} $  and  $E_{N+1}$ with respect to Fermi levels. In --S--(CH$_2)$$_N$--S-- wire, the terminal units  are attributed to either  sulfur's lone pair or the binding Au-S orbitals. Each of them   does  not enter in resonance with Fermi levels.  Therefore, the  terminal transmission functions (\ref{ltr0}) and (\ref{rtr0}) are monotonic in the integration region $\Delta E_{Ls} \geq \epsilon \geq \Delta E_{Rs}$. The similar  situation is true for terminal  NH$_2$ and  COOH units. This is due to the fact that in voltage region [-0.4,+0.4]V the orbital energies $E_{0} $  and  $E_{N+1}$  of these  units are positioned below chemical potentials  $\mu_L$ and $\mu_R$ (cf.  Fig. \ref{fig2}).  Thus,  bearing in mind  that  X--(CH$_2)$$_N$--X wires  contain same $N$ - alkane chains one can estimate  the conductance setting $g\simeq g^{(1)}$.

\begin{figure}
\includegraphics[width=7.3cm]{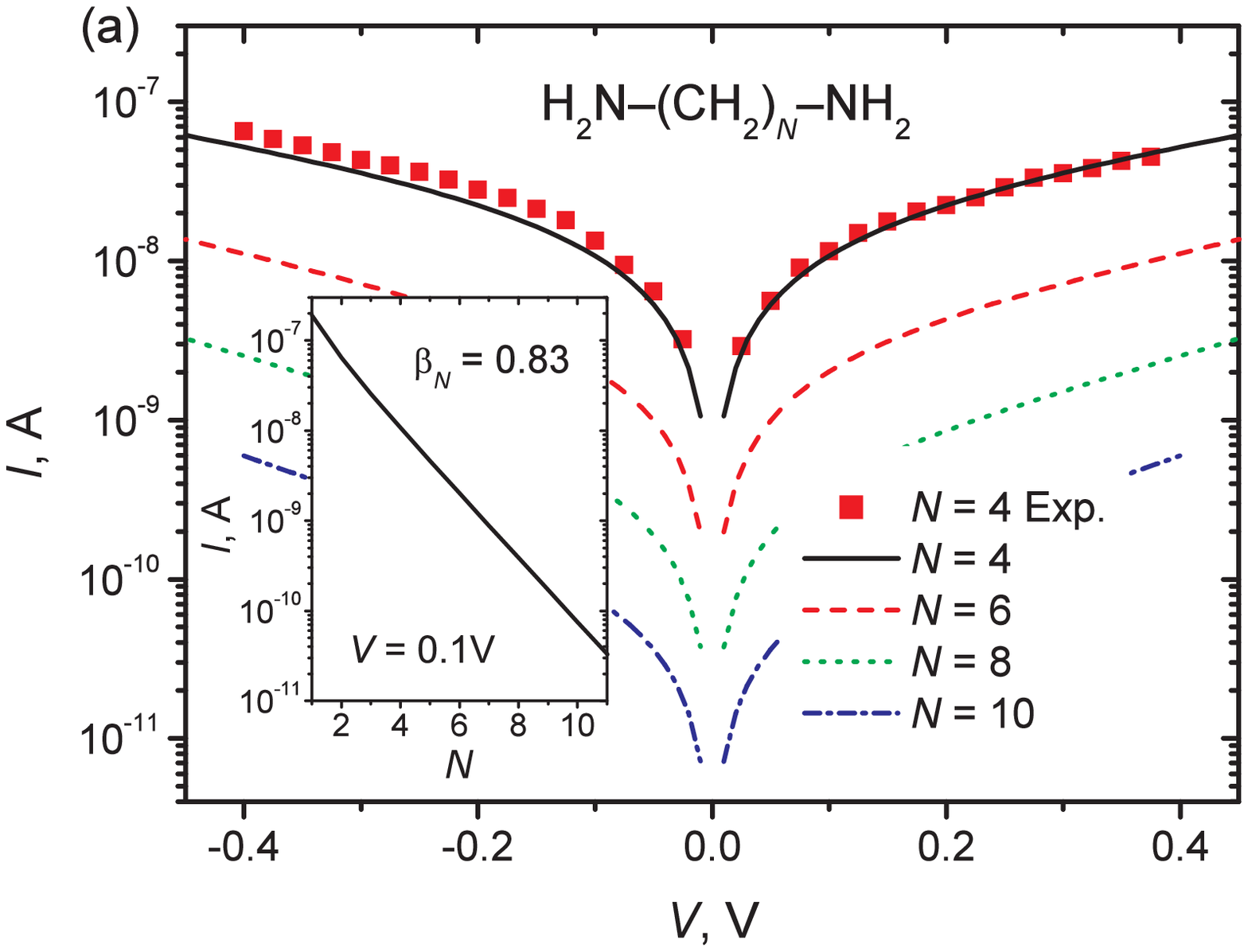}
\\

\includegraphics[width=7.3cm]{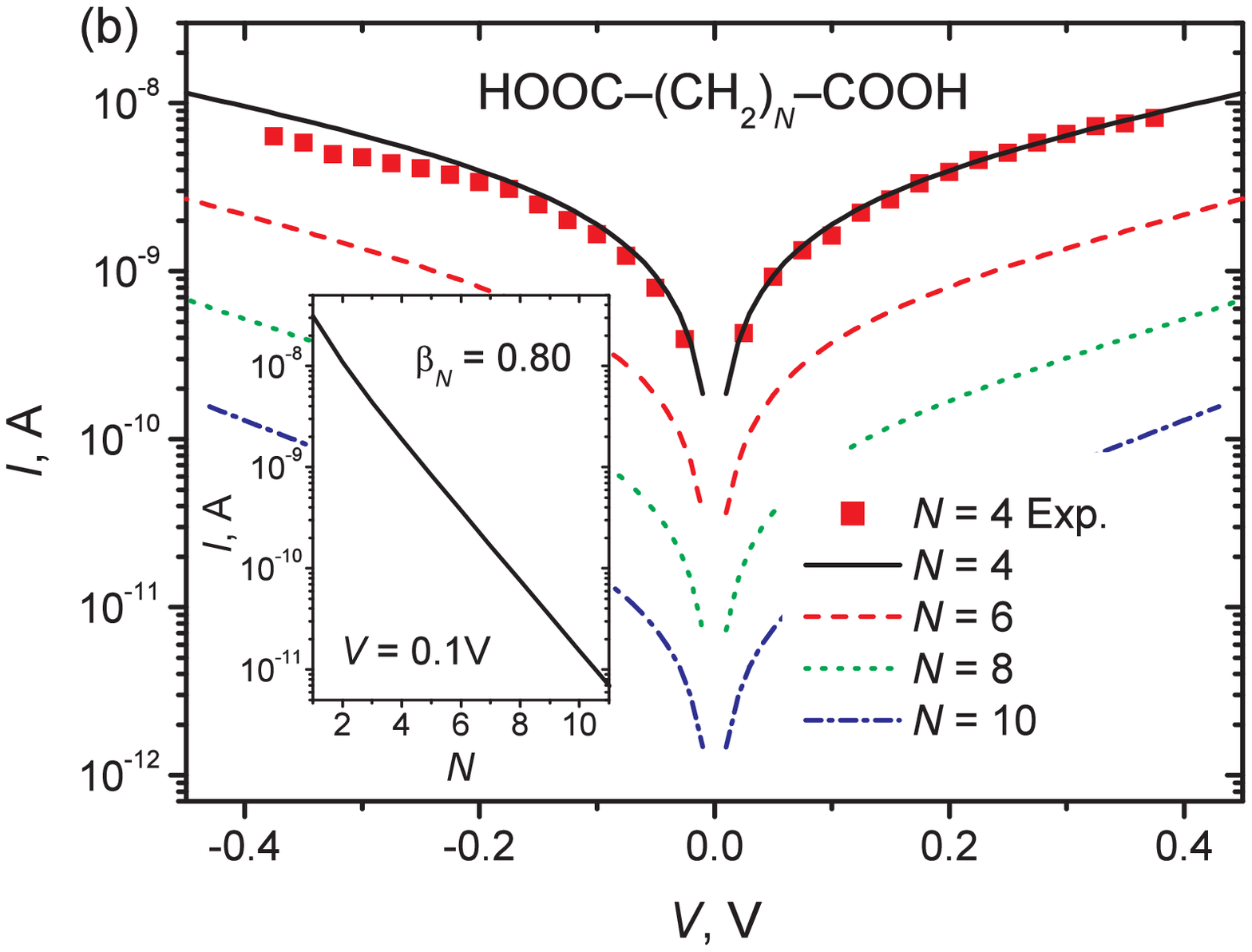}
\caption{$I/V$ characteristics of  $N$-alkanes terminated  with diamine (a) and dicarboxylic-acid anchoring groups. The data points represent the data adopted from experiment \protect\cite{fan06}.
Insertions show the exponential drop of nonresonant tunneling current at the fixed attenuation factor $\beta = \beta_N$ (symbol  $N$ indicates the drop per chain unit).
The curves are calculated with Eq. (\protect\ref{imvs}) at $N =4,6,8,10$. Calculation parameters are $\Delta E_*$ = 1.50 eV, $\Delta E_s^{(0)}$ = 5.72 eV, $t_*$ = 2.60 eV, $t_s $= 2.63 eV, $\Gamma_*$ = 0.30 eV  (a)  and $\Delta E_*$ = 1.25 eV,
$\Delta E_s^{(0)}$ = 6.03 eV, $t_*$ = 2.64 eV, $t_s$ = 2.79 eV, $\Gamma_*$ = 0.09 eV (b).
}
\label{fig4}
\end{figure}

\begin{figure}
\includegraphics[width=7.3cm]{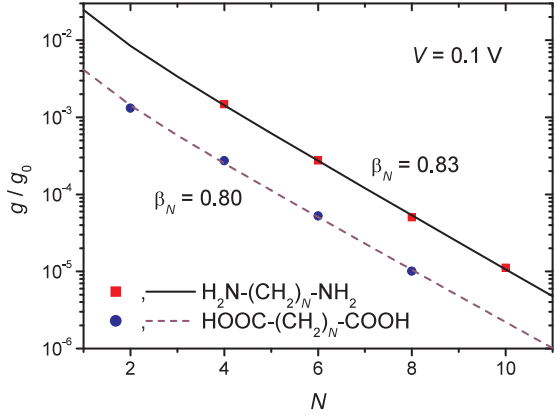}
\caption{Attenuation of the nonresonant  tunneling conductance with an increase of the number of C-C bonds of $N$ - alkanedithiol wire attached  to the electrodes through  NH$_2$ and COOH groups. The data points represent the data adopted from experiment \protect\cite{fan06}. Exponential approximation yields the same $\beta =\beta_N $  that of the current (see insertions in Fig.\protect\ref{fig4}).  Calculations of the $g \simeq g^{(1)}$
with use of Eq. (\protect\ref{g1b}). Parameters are the same as in Fig.\protect\ref{fig4}.
}
\label{fig5}
\end{figure}

To specify  fundamental parameters $\Delta E_s^{(0)}$ and $t_s$ for the wires  X--(CH$_2)$$_N$--X with X = NH$_2$, COOH,  one has to take into account the fact that different terminal units can change the  energy position of delocalized HOMO level with respect to the  Fermi levels \cite{fan06,kim06}. Thus,  the magnitudes of  $\Delta E_s^{(0)}$ and $t_s$ have to be nonidentical for different wires. For the wire with X = SH,  the attenuation factor $\beta_0$ exceeds the similar quantity for the wires with X = NH$_2$, COOH. Therefore, in line with basic equality (\ref{et}) one can assume that for the last two wires, the $\Delta E_s^{(0)}$  is smaller in value.
We estimate the $\Delta E_s^{(0)}$ comparing the theoretical expression for the current with the experimental $I/V$ characteristics at different number of CH$_{2}$ units. The theoretical description is based on the mean-value approximation for the  current, which   reads \cite{pet18}
%
\begin{equation}
I_{m.v.} = i_0\, |e|V \,
\frac{(\Gamma_*t_{*}^{2}/t_s)^2}{[\Delta E_{*}^2 - (|e|V
\eta_*)^2][\Delta E_{*}^2 - (|e|V)^2(1-\eta_*)^2
]}\overline{T}_{reg}(N)\,.
\label{imvs}
\end{equation}
Here, $\overline{T}_{reg}(N)$ being determined by Eqs. (\ref{att1}) -
(\ref{f}). Fig. \ref{fig4} shows a good fit of Eq.   (\ref{imvs}) to  the
data  if one sets  $\Delta E_s^{(0)}\approx 5.72$  eV,
$t_s\approx 2.63$ eV and  $\Delta E_s^{(0)}\approx 6.03$  eV,
$t_s\approx 2.79$ eV for  H$_2$N--(CH$_2)$$_N$--NH$_2$ and
HOOC--(CH$_2)$$_N$--COOH molecular wires, respectively.
With  the  same parameters,  we achieve the  fit  to the data for the conductance $g$ as a function of the  number of chain
units.   Fig. \ref{fig5}  depicts this behavior for  the ohmic regime.

Note, that for the description of the near-zero bias (z.b.) conductance, a much more simple form for the $g$ exists.
This follows directly from
Eq.  (\ref{g1b})  yielding
%
\begin{equation}
g = g_{z.b.} \simeq g_{unit} \Phi(\beta_0,N)
\label{nzbc}
\end{equation}
where
%
\begin{equation}
g_{unit} = g_0\Big(\frac{\Gamma_*}{\Delta E_{s}^{(0)}}\Big)^2
\Big(\frac{t_*}{\Delta E_{*}}\Big)^4
\label{nzbcu}
\end{equation}
is the conductance through a molecular wire with one bridging unit, and $\Phi(\beta_0,N)$ is the attenuation function (\ref{lf})
with attenuation factor $\beta_{0} = \beta(\epsilon =
\Delta E^{(0)}_{s})$. Form (\ref{nzbc}) refers to the wire  where terminal units  are coupled via the bridging units. This means  that  one can not set $N = 0$ in attenuation function $\Phi(\beta_0,N)$ to specify the contact conductance.

For the wire where   $\exp{[- (N+1)\beta (\epsilon)]}\ll 1$, the attenuation function  $\Phi(\beta (\epsilon),N)$  appears as
%
\begin{equation}
\Phi (\beta (\epsilon),N)  \approx \big(1 - {\rm e}^{-2\beta (\epsilon)}
\big)^2{\rm e}^{-\beta (\epsilon)(N - 1)}\,.                                                                                                                                                                                                                                                                                                                                                    \label{lfr}
\end{equation}
It demonstrates a pure  exponential drop with $N$.

At small inter-site coupling $t_s$, when condition
%
\begin{equation}
  (2t_s/\epsilon)^2 \ll1\,
\label{mcine}
\end{equation}
is satisfied, the attenuation factor (\ref{df}) is reduced to McConnel's form
%
\begin{equation}
  \beta (\epsilon)\simeq\beta_{M.C.} = 2\ln{(\epsilon/t_s)}\,.
\label{afmc}
\end{equation}
Another limiting case is realized if property
%
\begin{equation}
\Delta E \ll 2t_s\,
\label{inrbar}
\end{equation}
is satisfied  for quantity
 %
\begin{equation}
\Delta E  = \epsilon  -2t_s \,.
\label{barh}
\end{equation}
Physically, $\Delta E$ is the energy distance between transmission
energy $E$ and the position of the "the center of gravity"
(cf. Fig. \ref{fig2} and Eq. (\ref{shdel})). When the inequality (\ref{inrbar}) is satisfied, one can introduce the effective electron mass $m^*$ even though the regular chain  as a whole may be of  finite length \cite{pet18}.  The expression for the mass,
%
\begin{equation}
 m^*= \hbar^2/2t_sl_s^2\,,
\label{em}
\end{equation}
is determined by the intersite coupling $t_s$   and the distance $l_s$ between the neighboring sites of electron localization (cf. Fig. \ref{fig1}).  Introduction of the effective mass leads to the equality
%
\begin{equation}
\exp{[-\beta (\epsilon)(N - 1)]} = \exp{(-\beta_{B}\,d_s )}\
\label{supbar}
\end{equation}
where
%
\begin{equation}
\beta_{B} =  (2/\hbar) \sqrt{2m^*\Delta E}
\label{afrb}
\end{equation}
is the attenuation factor (in \AA$^{-1}$)  and
$d_s = l_s(N-1)$ is the distance (in \AA) between the edge chain  units $n=1$  and $n=N$. It follows from Eq. (\ref{afrb}) that the $\Delta E$ can be referred to as the height of rectangular barrier of
length $d_s$. Note, however, that such an interpretation can be used only if specific condition (\ref{inrbar}) exists during a transmission process.

Since near-zero bias tunneling occurs at $E \approx E_F$, then $\epsilon \approx \Delta E_s^{(0)}$. This means that $\Delta E_0 = \Delta E_s^{(0)} - 2t_s \approx $ 0.46 eV and 0.45 eV for H$_2$N--(CH$_2)$$_N$--NH$_2$ and
HOOC--(CH$_2)$$_N$--COOH molecular wires, respectively.
As far as the gap $\Delta E_0$ satisfies the condition (\ref{inrbar}), it becomes possible to interpret a nonresonant  superexchange transmission across
H$_2$N(HOOC)--(CH$_2)$$_N$--NH$_2$(COOH) molecular
wire  as a tunneling of an electron with an effective mass
$m^* = 0.85 (0.80) m_e$  through a rectangular barrier of  the height $\Delta E_0$ = 0.46 (0.45) eV and the length $d_s = l_s(N-1)$,($m_e$ is the elementary electron mass). In line with relation(\ref{supbar}), the corresponding zero-bias barrier attenuation factor (\ref{afrb})
reads $\beta_B^{(0)} = \beta_0l_s^{-1}$.
In the biased LWR system, the conductance drop with chain length is determined by the attenuation functions  $\Phi(\beta_L,N)$ and $\Phi(\beta_R,N)$.  Similarly,  with the near-zero bias case, one can express  the chain attenuation
factor $\beta_r$, ($r =L,R$), via  the barrier one,
%
\begin{equation}
\beta_r = l_s \beta_B^{(r)}
\label{lrsb}
\end{equation}
where
%
\begin{equation}
\beta_B^{(r)} = (2/\hbar)\sqrt{2m^*\Delta E_{r}}\,.
\label{attlr}
\end{equation}
In accordance with  Eq. (\ref{xes}) the dependence of the barrier height on the bias voltage reads
%
\begin{equation}
\Delta E_{r} =  \Delta E_0 + (|e|V/2) (\delta_{r,L} - \delta_{r,R})\,.
\label{attlr0}
\end{equation}
Thus, in the $N - $alkanes terminated by SH, NH4$_{2}$
and COOH anchoring groups,  the analysis of  the conductance
drop can be also  performed using the barrier
model, until inequalities (\ref{inedel}) and (\ref{inedel}) are satisfied at the tunneling transmission.
In a barrier model, the  height, Eq. (\ref{barh}) and
the tunneling effective electron mass, Eq. (\ref{em}) are expressed via the characteristics of a molecular junction.  This is  reflected in relation  (\ref{lrsb}) between respective attenuation factors. The relation exists until
the inequality
%
\begin{equation}
\Delta E_{L(R)} \ll 2t_s\,
\label{inrbarr}
\end{equation}
retains its validity during the tunneling charge transmission. However, it is necessary to note that apart from attenuation factors, Eq. (\ref{lrsb}), additional quantities exist that specify the current and the conductance.For a perfectly symmetric  LWR system, they are the following:
coupling of the terminal unit to the corresponding chain edge
unit ($t_*$), MO's broadening ($\Gamma_*$), the voltage
division factor ($\eta_*$),  and the terminal gap ($\Delta E_*$).
In the case of an ohmic regime, some of these quantities are
combined in a single parameter $g_{unit}$, Eq. (\ref{nzbcu})
characterizing the conductance of an elementary LWR system with the single bridging unit.  As  follows from Fig. \ref{fig4},
$g_{unit}$ is about $ 2.5\cdot 10^{-2} g_0$ and
$ 4\cdot 10^{-3} g_0$
for H$_2$N--(CH$_2)$$_N$--NH$_2$ and
HOOC--(CH$_2)$$_N$--COOH molecular wires, respectively.
The $g_{unit}$ does not contain the barrier
characteristics of a regular chain.  Thus, even though the rigorous correspondence, Eq.  (\ref{lrsb}) exists between the attenuation factors, it is  more preferable  to  explain the physics of a tunneling process in the framework of superexchange model.

At given $V$ and $N$, the feasibility of the modified superexchange model is limited by inequality (\ref{inedel}). For $N - $alkanes with terminal NH$_2$ and COOH groups where $t_s \approx 2.7$ eV, the model works at $N\leq 20$  and $N\leq 10$ if  if $V =0.1$ V and V =0.4 V, respectively. Therefore, for instance, the theoretical curves in Fig.  \ref{fig4} that cover the experimental data at $N=4$, predict
$I/V$ characteristics at $N =6,8,10$.

\section{Conclusions}

The main objective of this study was to obtain explicit expressions for the tunneling conductance in  a molecular wire consisting of a regular chain connected to metal electrodes through terminal groups or individual atoms. In the case of a nonresonant electron/hole transmission through such a molecular junction, a temperature independent current and conductance were observed, and their
values  decay exponentially with increasing length of the wire's interior range (regular chain).
Since the exponential attenuation indicates the tunneling nature of the conductivity in the "electrode-molecular wire-electrode" system, the analysis of experimental
current-voltage characteristics and the conductance in the LWR system is most often carried out using the Simmons barrier model by analogy with the "electrode-dielectric-electrode" structures. However, this  phenomenological model does not sufficiently  reflects  the specifics of a current/conductance formation in molecular junctions. The progress in understanding the mechanism of conductivity in the molecular wires is due to McConnel's model of distant superexchange transfer of electrons / holes. In this model,  the current decays exponentially similar  the barrier model. At the same time, McConnel's model has restrictions related to the
applicability of the perturbation theory in the parameter that characterizes ratio (\ref{mcine}) between the inter-site coupling $t_s$ and  transmission gap $\epsilon$.

In the present paper, the modified model of superexchange tunneling in the version proposed in ref. \cite{pet18} is  used  for the description of the nonresonant current through a molecular wire. The model assumes a much more soft relation,  Eq. (\ref{ets}) between quantities  $t_s$ and
$\epsilon$ as in comparison  with the model  of deep tunneling. This allows one to derive the attenuation factor $\beta(\epsilon)$, Eq. (\ref{df})  that, in
limiting cases, is reduced to that for  McConnel's, Eq. (\ref{afmc}) or barrier's, Eq.    (\ref{afrb}) models. It is shown that when analyzing the current through molecular wires, the  rectangular barrier model can work only in the case of a strong delocalization of an electron/hole in virtual states of  a regular chain. It is shown that the delocalization is conserved only if the condition (\ref{inedel})
is satisfied at given $V$ and $N$. Besides, the height of apparent zero-bias rectangular barrier $\Delta E$ has to be much less than the doubled parameter of site-site coupling (cf. Eq (\ref{inrbar})). In the case of superexchange tunneling mediated by the virtual chain HOMOs, the zero-bias barrier coincides with the energy gap
between the Fermi level and the delocalized HOMO level of a
long chain. At nonzero bias voltage, this barrier is
transformed into  two, Eq. (\ref{attlr0}) that have to
satisfy the condition (\ref{inrbarr}). It is important to note that even though a rigorous correspondence is established between the attenuation factors (cf. Eq. (\ref{lrsb})), a more complete description of the wire conductance occurs with use of superexchange model.
This is due to the  superexchange model
allowing one to obtain not only the chain  attenuation functions $\Phi(\beta_L,N)$  and $\Phi(\beta_R,N)$, but also the factors that specify properties of the elementary molecular wire with a single bridging unit.  In Eq. (\ref{g1b}), these basic superexchange  factors are presented just before functions $\Phi(\beta_L,N)$  and $\Phi(\beta_R,N)$. In the ohmic regime of the tunneling transmission, these superexchange   factors reduce to  the  $g_{unit}$, Eq. (\ref{nzbcu}).

The analysis of the nonresonance tunneling conductance  in --(CH$_2)$$_N$--NH$_2$ and HOOC--(CH$_2)$$_N$--COOH molecular wires shows that the modified superexchange model is quite appropriate to explain the experimental results. In framework of the model, a formation of the conductance is associated with the virtual participation of the localized HOMOs of terminal units H$_2$N or COOH as well as the delocalized chain HOMOs formed from the localized C-C bonds. In the case of ohmic transmission regime, the attenuation of the conductance with length of $N-$alkane chain may  be interpreted as the process of electron tunneling through a rectangular barrier. The barrier height and width as well as  the effective mass of the tunneling electron are determined via the characteristics of the  $N-$alkane chain. However, such interpretation is possible only for those $V$ and $N$ at which principal inequalities (\ref{inedel}) and  (\ref{inrbar}) are satisfied.

Present study shows, that  the analysis of  current and conductance characteristics with use of the modified superexchange model opens new possibilities of understanding the mechanism  of tunneling charge transfer processes in linear molecular junctions.

\section {Acknowledgments}
E.G.P. and E.V.S. acknowledge the support by the NAS Ukraine via project 0116U002067. V.S. acknowledges the support of the MSU Global Education Grant. This project has received funding from the European Union's Horizon 2020 research and innovation programme under the Marie Sklodowska-Curie grant agreement  NANOGUARD2AR 690968 (S.L)  and ENGIMA No. 778072 (V.V.G. and A.V.R.).


%

\begin{thebibliography}{00}
\bibitem{foot1} At strong interaction between the edge atom of the wire and the electrode's surface atoms, the coupling between the terminal unit and the electrode has to be associated with the  bind of the edge atom with the surface atoms. For instance, if S is the end atom, then the single - bond coupling has to be reffered to the S-Au bind and the nearest  surface Au atoms.
\end{thebibliography}
\end{document}